\documentclass[5p]{elsarticle}
\pdfoutput=1
\usepackage{amsmath,amssymb,amsfonts,amsbsy}
\usepackage{physics}
\usepackage[disable]{todonotes}
\usepackage[bookmarks=true,colorlinks=true,linkcolor=black,citecolor=black,urlcolor=black,bookmarksnumbered]{hyperref}

\newcommand{\ad}{\operatorname{ad}}
\newcommand{\Ad}{\operatorname{Ad}}

\newcommand{\ads}{\mathrm{AdS}_5 \times \mathrm{S}^5}

\newcommand{\curvJ}{\mathcal{F}(I)}
\newcommand{\curvL}{\mathcal{F}(L)}

\title{Yang-Baxter deformations of the flat space string
}
\author{Khalil Idiab}
\ead{khalil.idiab@physik.hu-berlin.de}
\author{Stijn J. van Tongeren}
\ead{svantongeren@physik.hu-berlin.de}
\address{Institut f\"ur Mathematik und Institut f\"ur Physik, Humboldt-Universit\"at zu Berlin, \\ IRIS Geb\"aude, Zum Grossen Windkanal 2, 12489 Berlin, Germany}

\begin{document}

\begin{abstract}
  We define integrability preserving Yang-Baxter deformations of symmetric space sigma models with non-semi-simple symmetry group, in particular the flat space string, using only the essential structures of a symmetric space sigma model. For homogeneous deformations, the Lax connection is of the same form as the semi-simple case, although the $R$ operator no longer satisfies a freestanding operator equation. For inhomogeneous deformations, the form of the Lax connection needs to be relaxed, by modifying the underlying algebra. We illustrate the construction by discussing nonabelian deformations of three dimensional Minkowski space.
\end{abstract}


\maketitle

\section{Introduction}\label{sec:intro}

Yang-Baxter deformations of integrable sigma models have attracted considerable attention in recent years, in particular given their applications in string theory and AdS/CFT. These deformations were originally defined for the principal chiral model on a simple group \cite{Klimcik:2002zj,Klimcik:2008eq}, and involved a solution of the inhomogeneous classical Yang-Baxter equation. Yang-Baxter deformations were subsequently extended to symmetric space \cite{Delduc:2013fga} and semi-symmetric space sigma models \cite{Delduc:2013qra}, and to solutions of the homogeneous classical Yang-Baxter equation \cite{Kawaguchi:2014qwa}. These models all assume a (semi-)simple group or coset denominator, thereby guaranteeing the existence of a nondegenerate bilinear form (the Killing form), whose full adjoint invariance is used in the construction. In this paper we will let go of this restriction, and consider Yang-Baxter deformations of non-semi-simple symmetric space sigma models. This makes it possible to investigate Yang-Baxter deformations of the flat space string, with non-semi-simple Poincar\'e symmetry, intended to function as a simpler analogue of deformations of e.g. the $\ads$ string.

The original Yang-Baxter deformation of the $\ads$ superstring \cite{Delduc:2013qra,Delduc:2014kha} is based on the standard inhomogeneous solution of the classical Yang-Baxter equation. Due to the noncompact nature of $\ads$ and the presence of fermions -- algebraically, the choice of real form and of Dynkin diagram, respectively -- it is possible to define further inhomogeneous deformations differing already in terms of the bosonic geometry \cite{Delduc:2014kha}, or only in terms of the fermions \cite{Hoare:2018ngg}. Out of these various options, only deformations based on the fully fermionic Dynkin diagram are what is known as unimodular, meaning they are guaranteed to give one-loop Weyl invariant superstrings \cite{Borsato:2016ose}, see also \cite{Hronek:2020skb}. At the quantum level the various possible inhomogeneous deformations lead to a trigonometric $q$ deformation of the exact S matrix of the $\ads$ string \cite{Beisert:2008tw,Hoare:2011wr,Arutyunov:2013ega,Seibold:2020ywq}, and can be studied using Bethe ansatz techniques \cite{Arutynov:2014ota,Seibold:2021rml}. From this point of view, despite their (super)geometric differences, these various inhomogeneous deformations appear to be equivalent \cite{Hoare:2016ibq,Seibold:2020ywq,Seibold:2021rml}.

Homogeneous deformations of the $\ads$ superstring \cite{Kawaguchi:2014qwa} can instead take various fundamentally inequivalent forms, giving known and new string backgrounds, see e.g. \cite{Matsumoto:2015jja,Matsumoto:2015uja,vanTongeren:2015soa,Kameyama:2015ufa,Kyono:2016jqy,Hoare:2016hwh,Osten:2016dvf,vanTongeren:2016eeb}. These models correspond to a type of twisting of the undeformed string \cite{Vicedo:2015pna,vanTongeren:2015uha,vanTongeren:2018vpb,Borsato:2021fuy}, and are conjectured to give string duals to noncommutative version of maximally supersymmetric Yang-Mills theory \cite{vanTongeren:2015uha}. In this class of models, only abelian deformations built out of the Cartan generators of the symmetry of the string, readily admit a quantum description in terms of the Bethe ansatz \cite{vanTongeren:2021jhh}. Other types of homogeneous deformations, be it almost abelian, jordanian, or otherwise nonabelian, necessarily break the light-cone isometries used in the typical exact S matrix and Bethe ansatz approach to the $\ads$ string. The best current understanding of their structure is in terms of the classical spectral curve, recently worked out in \cite{Borsato:2021fuy}. Understanding these models at the quantum level is an important open challenge.\footnote{Very recently, first steps have been taken in the systematic quantization of integrable models based on an affine Gaudin model formulation \cite{Kotousov:2022azm}, which in principle could be applied here as well.}

One way to approach the quantum structure of ``non-Cartan'' homogeneous deformations of the $\ads$ string, is to find and study a simpler analogue, i.e. investigate similar deformations for the flat space string. Unfortunately, the conventional construction of the Yang-Baxter model does not directly apply to the flat space string. This is due to the lack of a nondegenerate but fully adjoint invariant bilinear form: the Poincar\'e algebra only admits a Lorentz but not Poincar\'e invariant nondegenerate bilinear form. Because of this, only particular deformations of the flat space string have been investigated thus far, where it is possible to avoid this obstacle.

One approach is essentially to embed the $d$-dimensional Poincar\'e algebra in $\mathfrak{so}(2,d)$, which is simple, and build Yang-Baxter deformations on this auxiliary structure. This was pursued in four dimensions in \cite{Matsumoto:2015ypa,Borowiec:2015wua}. Another approach is to consider the flat space string as the flat space limit of $\ads$, and consider the same flat space limit for deformations of the $\ads$ string \cite{Pachol:2015mfa,Hoare:2016hwh}.\footnote{For inhomogeneous deformations this presents an interesting link to the mirror model of the $\ads$ string \cite{Arutyunov:2014cra,Arutyunov:2014jfa,Pachol:2015mfa}.} Both these approaches have limitations however. The former embedding approach is a rather ad hoc construction, which as we will show is unnecessary, and its extension to the flat space superstring is not clear. Conversely, the latter flat space limit approach can only describe the subset of deformations that arise as limits of deformations of $\ads$ or similar spaces, and moreover appears unnecessarily cumbersome if we are interested in the flat space string itself. As we will show, it is possible to define the action of Yang-Baxter deformations of the flat space string purely in terms of Poincar\'e-algebraic structures. In our approach, at the level of the Lax connection, inhomogeneous deformations do require an algebra beyond Poincar\'e, namely $\mathfrak{so}(2,d-1)$ or $\mathfrak{so}(1,d)$.

In this paper we will take a general approach, and define Yang-Baxter deformations for symmetric space sigma models with non-semi-simple symmetry group. In practice we focus on sigma models on flat space, in particular with Lorentzian signature, but most of our discussion applies generally. We start by reviewing the essential ingredients for a symmetric space $G/H$ sigma model, and how they suffice to give a Lax connection for the undeformed model. We then consider the standard Yang-Baxter deformation, and discuss how the conventional definition of the $R_g$ operator in the action needs to be modified in order to give meaningful deformations. Instead of composing a constant $R$ operator with adjoint actions of the group element $g$, we simply define the $R$ operator via an adjointly transformed $r$ matrix, a natural algebraic object. In the semi-simple setting, this definition is equivalent to the standard one. To show that this is an integrable deformation, we find a Lax representation for the resulting equations of motion, mimicking the derivation in the semi-simple case. The difference is that the $R_g$ operator does not satisfy the classical Yang-Baxter equation in the operator form typical of the semi-simple setting, but rather an analogous equation that follows directly from the standard CYBE for the $r$ matrix without using adjoint invariance. We illustrate the kind of deformations this generates, by discussing various nonabelian deformation of $\mathbb{R}^{1,2}$.


\section{The flat space coset sigma model}

It is well known that the string sigma model on Minkowski space can be represented by a symmetric space sigma model. We will briefly review its construction and establish our conventions, before discussing its deformations.

A general symmetric space can be described as a coset $G/H$ where the Lie algebra $\mathfrak{g}$ of $G$ admits a $\mathbb{Z}_2$ grading
\begin{align}
\label{eq:7}
\mathfrak{g}=\mathfrak{g}^{(0)} \oplus \mathfrak{g}^{(1)},&& [\mathfrak{g}^{(i)},\mathfrak{g}^{(j)}]\subset \mathfrak{g}^{(i+j\hspace{-0.15cm}\mod 2)}.
\end{align}
such that $\mathfrak{g}^{(0)} =\mathfrak{h}$ is the Lie algebra of $H$.

We want to equip $\mathfrak{g}$ with a symmetric bilinear form, $\ip{\bullet}:\mathfrak{g} \times \mathfrak{g} \rightarrow \mathbb{R}$, that is grade compatible,
\begin{equation}
\label{eq:gradecompatibility}
\ip{X^{(0)}}{Y^{(1)}}=0, \qquad \forall\, X,Y \in \mathfrak{g},
\end{equation}
$\Ad_{H}$ invariant,
\begin{equation}
\label{eq:AdHinvariance}
\ip{X}{Y}=\ip{h^{-1}X h}{h^{-1} Y h}, \qquad \forall\, h \in H,\, X,Y \in \mathfrak{g},
\end{equation}
equivalently,
\begin{equation}
\ip{[Z,X]}{Y}+\ip{X}{[Z,Y]}=0, \qquad \forall\, Z \in \mathfrak{h},\, X,Y \in \mathfrak{g},
\end{equation}
and nondegenerate at least on $\mathfrak{g}^{(1)}$.\footnote{Strictly speaking the action only requires a form on $\mathfrak{g}^{(1)}$, but it is useful to extend this to a form on all of $\mathfrak{g}$.} At times we will use an abstract basis $\{T_i\}$ $i=1,\ldots,\mbox{dim}(\mathfrak{g})$ for $\mathfrak{g}$, with structure constants $[T_i,T_j] = f_{ij}{}^{k}T_k$. Since our bilinear form is nondegenerate by assumption, we will use $L_{ij} \equiv \ip{T_i}{T_j}$ and its inverse to lower and raise indices respectively. When $G$ is semi-simple we can take the Killing form as our bilinear form, which automatically has a larger $\mathrm{Ad}_G$ invariance.

To construct an action we now consider a $G/H$-valued field $g$ living on a two dimensional worldsheet $\Sigma$: $g:\Sigma\rightarrow G/H $, and the associated Maurer-Cartan form $A_{\alpha}=-g^{-1}\partial_{\alpha}g \in \mathfrak{g}$. In terms of these, the sigma model action is
\begin{align}
\label{eq:undefAction}
S[\gamma,g]=\frac{1}{2}\int_{\Sigma}d^{2}\sigma \gamma^{\alpha\beta}\ip{A_{\alpha}}{\mathcal{P}A_{\beta}},
\end{align}\todo{prefactor? citation?}
where $\mathcal{P}:\mathfrak{g}\rightarrow\mathfrak{g}^{(1)}$ is projector, $\gamma^{\alpha\beta}=\sqrt{h}h^{\alpha\beta}$, where $h_{\alpha\beta}$ is the worldsheet metric, and $\sigma^\alpha$ are the worldsheet coordinates. For $g$ to live in the coset, the action \eqref{eq:undefAction} must have gauge invariance under local right $H$ transformations $g\rightarrow gh$, $h\in H$. This gauge invariance follows directly from the grading of the algebra, combined with the grade compatibility and $\Ad_{H}$ invariance of the bilinear form. Taking a particular set of coordinates on $G/H$ can be done by choosing a coset representative $g= g(x)$. This gives the coordinate action
\begin{equation}
\label{eq:coordinateaction}
 S=\int_{\Sigma}d^{2}\sigma P_{-}^{\alpha\beta} (G_{\mu\nu}+B_{\mu\nu}) \partial_{\alpha}x^{\mu}\partial_{\beta}x^{\nu},
\end{equation}
where we have included a $B$ field for future reference, and  $P_{\pm}^{\alpha\beta}= \frac12 \left(  \gamma^{\alpha\beta}\pm \epsilon^{\alpha\beta}\right)$. Below we typically denote the action of these world sheet projection operators $P_\pm$ by a $\pm$ subscript, i.e. $P_\pm^{\alpha\beta} Y_\beta = Y^\alpha_\pm$, with associated identities like $X_\alpha Y^\alpha_\pm = X_{\mp,\alpha} Y^\alpha_\pm$.

Taking a coset with $G=\mathrm{ISO}(1,d-1)$ and $H=\mathrm{SO}(1,d-1)$ gives $d$ dimensional Minkowski space, $G/H \simeq \mathbb{R}^{1,d-1}$. Parametrizing the Lorentz algebra $\mathfrak{h}$ by the standard generators $m_{\mu\nu}$ with commutation relations
\begin{equation}\label{eq:lorentzcommutators}
[m_{\mu\nu},m_{\rho\sigma}]=\eta_{\mu\sigma} m_{\nu\rho}+\eta_{\nu\rho}m_{\mu\sigma}-\eta_{\mu\rho}m_{\nu\sigma}-\eta_{\nu\sigma}m_{\mu\rho},
\end{equation}
where $\eta_{\mu\nu}$ is the Minkowski metric, and the remaining $\mathfrak{g}^{(1)}$ by the translation generators $p^\mu$ with
\begin{equation}\label{eq:mpcommutators}
[m_{\mu\nu},p_{\rho}]=\eta_{\nu\rho} p_{\mu}-\eta_{\mu\rho}p_{\nu}, \qquad [p_{\mu},p_{\nu}]=0,
\end{equation}
illustrates the $\mathbb{Z}_2$ graded structure of the Poincar\'{e} algebra. The desired Lorentz invariant nondegenerate bilinear form is given by
\begin{equation}
\label{eq:bilinearform}
\begin{aligned}
  \ip{m_{\mu\nu}}{m_{\rho\sigma}} & = \eta_{\mu\sigma}\eta_{\nu\rho}-\eta_{\mu\rho}\eta_{\nu\sigma},\\
  \ip{m_{\mu\nu}}{p_{\rho}}& =0, \qquad \ip{p_{\mu}}{p_{\nu}}=\eta_{\mu\nu}.
\end{aligned}
\end{equation}
If we now take the coset representative $g = e^{x_\mu p^\mu}$, we find the coordinate action \eqref{eq:coordinateaction} with
\begin{equation}\label{eq:undeformedmetricandbfield}
G_{\mu\nu} = \eta_{\mu\nu}, \qquad B_{\mu\nu}=0,
\end{equation}
as expected. The construction for $\mathbb{R}^{n,d-n}$ with different signature here and below is the same, with suitably changed metric.

\subsection{Equations of motion}

To find the equations of motion associated to the symmetric space sigma model, we consider the variation of the action under variations of $g$
\begin{align}
\delta_{g}S[\gamma,g]=\int_{\Sigma}d^{2}\sigma \gamma^{\alpha\beta}\ip{\delta_g A_{\alpha}}{\mathcal{P}A_{\beta}},
\end{align}
and note that the variation of the Maurer-Cartan form $\delta_g A_{\alpha}=[A_{\alpha} ,\epsilon]  -\partial_{\alpha}\epsilon,$ with $\epsilon=g^{-1}\delta g \in \mathfrak{g}$. Integrating by parts and dropping the boundary term, this gives
\begin{align}
\label{eq:undefVaried}
\delta_{g} S[\gamma,g]&= \int_{\Sigma} d^{2}\sigma \Big[\braket{\epsilon}{ \partial_{\alpha}\mathcal{P} A^{\alpha} }+ \braket{[A_{\alpha} ,\epsilon]}{\mathcal{P} A^{\alpha}}\Big].
\end{align}
We now need to bring the Lagrangian density to the form $\mathcal{L}=\ip{\epsilon}{\mathcal{E}}$, in order to use nondegeneracy of the form to identify $\mathcal{E}$ as the equations of motion. The second term of \eqref{eq:undefVaried} can trivially be brought to this form if the bilinear form is $\mathrm{Ad}_G$ invariant, but we only assume $\mathrm{Ad}_H$ invariance. Fortunately, due to the symmetric space structure the result is the same. Namely, since the bracket does not contribute for elements with non-matching grading, we have
\begin{align}
\ip{\left[A_{\alpha} ,\epsilon\right]}{ \mathcal{P} A^{\alpha}}&= \ip{\big[A^{(0)}_{\alpha} ,\epsilon^{(1)}\big]}{ \mathcal{P} A^{\alpha}}+\braket{\big[A_{\alpha}^{(1)} ,\epsilon^{(0)}\big]}{\mathcal{P} A^{\alpha}},\nonumber\\
  &=- \gamma^{\alpha\beta} \braket{\epsilon}{\big[A_{\alpha}^{(0)} , A^{(1)}_{\beta}\big]}
\end{align}
where in the second equality we used $\mathrm{Ad}_H$ invariance to move the grade zero terms to the other side, where the second term vanishes since $\gamma^{\alpha\beta}$ is symmetric. We hence find the equations of motion
\begin{align}
\label{eq:undeformedEOM}
\mathcal{E}=\partial_{\alpha} \mathcal{P} A^{\alpha}-\left[A_{\alpha} , \mathcal{P} A^{\alpha}\right]=0.
\end{align}
These equations are well known to admit a Lax representation \cite{Eichenherr:1979ci}, of the form
\begin{align}
\label{eq:undeflax}
L=A^{(0)}+\ell_{-}A_{-}^{(1)}+\ell_{+}A_{+}^{(1)},
\end{align}
where for a generic symmetric space we have the constraint that $\ell_{+}\ell_{-}=1$, solved by e.g. $\ell_{\pm}=z^{\pm 1}$ with $z$ as the spectral parameter. Specifically for the Poincare algebra where $[\mathfrak{g}^{(1)},\mathfrak{g}^{(1)}]=0$, and hence $A^{(0)}$ is flat on its own, the only constraint on $\ell_{\pm}$ is that $\ell_{+}\neq \ell_{-}$, allowing for a wider class of Lax connections.

\section{Yang-Baxter deformations}

The symmetric space sigma model with semi-simple $G$ is known to admit integrability preserving deformations, known as Yang-Baxter deformations.\footnote{Semi-simplicity is typically assumed, but strictly speaking the basic construction applies to any $G$ with a quadratic Lie algebra.} The corresponding action is \cite{Delduc:2013fga,Delduc:2013qra}
\begin{align}
\label{eq:YBaction}
  S[\gamma,g]&=\int_{\Sigma}d^{2}\sigma P_{-}^{\alpha\beta}\ip{\mathcal{P}A_{\alpha}}{D A_{\beta}},\\  D&=\frac{1}{1-\eta R_{g}\mathcal{P}}
\end{align}
where $R_{g}=\Ad_{g}^{-1}\circ R\circ \Ad_{g}$ with $R$ a constant linear map $R: \mathfrak{g}\rightarrow\mathfrak{g}$. The undeformed model is obtained at $\eta=0$. This deformation preserves integrability provided $R$ is antisymmetric, $\ip{A}{RB} = \ip{R^T A}{B}= -\ip{R A}{B}$, and satisfies the classical Yang-Baxter equation (CYBE) in operator form
\begin{equation}
\label{eq:operatorCYBE}
[R(X),R(Y)] - R([R(X),Y]+[X,R(Y)])= -c^2 [X,Y],
\end{equation}
where $c=1,i,0$, corresponding to inhomogeneous nonsplit, inhomogeneous split, and homogeneous deformations respectively. This form of the CYBE is equivalent to the standard CYBE\footnote{Here $r \in \mathfrak{g} \otimes \mathfrak{g}$, and $r_{ij}$ denotes $r$ embedded nontrivially in spaces $i$ and $j$ in $\mathfrak{g}^{\otimes3}$}
\begin{equation}
\label{eq:matrixCYBE}
 [[r,r]] \equiv [r_{12}, r_{13}]+[r_{12},r_{23}]+[r_{13},{r_{23}}] = -c^2 \Omega,
\end{equation}
where $\Omega \in \Lambda^3 (\mathfrak{g})$ is the canonical triple tensor Casimir of $G$, for semi-simple groups $\Omega = f^{ijk} T_i \wedge T_j \wedge T_k$.\footnote{Here the wedges denote a totally antisymmetric tensor product, including a $1/n!$ for an $n$-fold product. Invariance of $\Omega$ is equivalent to the Jacobi identity for the structure constants.} If we express $R$ in terms of $r$ via
\begin{equation}
\label{eq:Rnog}
R(X) = \ip{r}{X}_1,
\end{equation}
where the subscript indicates the tensor factor where the bilinear form is applied, the operator form of the CYBE is just $\ip{[[r,r]]+c^2 \Omega}{X,Y}_{1,2} = 0$, upon using the $\mathrm{Ad}_G$ invariance of the bilinear form for semi-simple $G$.

In our non-semi-simple setting with an only $\mathrm{Ad}_H$ invariant bilinear form, the above definitions do not give a suitable deformation -- it is for instance possible to insert nontrivial $R$ operators and get no deformation.\footnote{Take e.g. four dimensional Minkowski space with coset representative $g = e^{x^\mu p_\mu}$ and deform by $r = m_{01}\wedge m_{23}$. Then $A = dx^\mu p_\mu = \mathcal{P}(A)$ and $R_g(\mathcal{P}(A)) = g^{-1} R(dx^\mu p_\mu) g = 0$. The action is then undeformed, contrary to general expectations.} The expected deformation of, for instance, the background geometry, can fortunately be readily restored by moving an $\Ad_{g}$, i.e. by defining the inherently $g$-dependent linear operator $R_g$ as
\begin{equation}
\label{eq:Rgnonsemisimple}
R_{g}(X) = r^{ij}\ip{  g^{-1}T_{i}g}{X} g^{-1}T_{j}g,
\end{equation}
where we use our basis $\{T_i\}$ for $\mathfrak{g}$. Equivalently
\begin{equation}
\label{eq:Rgviarg}
R_g(X) = \ip{r_g}{X}_1,
\end{equation}
where $r_g = \mathrm{Ad}^{-1}_g \otimes \mathrm{Ad}^{-1}_g (r)$ with $r = r^{ij} T_i \otimes T_j$. These definitions of $r_g$ and $R_g$ are algebraically natural, and equivalent to the semi-simple ones in that setting.\footnote{Both definitions for $R_g$ preserve gauge invariance of the action also in our non-semi-simple setting, but only the second gives the desired deformations.} We now want to show that the equations of motion for the deformed action \eqref{eq:YBaction}, with $R_g$ as in equations (\ref{eq:Rgnonsemisimple}, \ref{eq:Rgviarg}), admit a Lax representation.

\subsection{Equations of motion}

Similarly to the undeformed model, the combination of the symmetric space grading and the $H$-adjoint invariance is sufficient to cast the equations of motion in a suitable form, one familiar from the semi-simple case \cite{Delduc:2013fga,Delduc:2013qra}. The variation of action with respect to $g$ gives three terms
\begin{align}
\label{eq:13}\nonumber
  \delta_{g} S[\gamma,g]=\int d^{2}\sigma P_{-}^{\alpha\beta}  \big( \braket{ \delta A_{\alpha}}{\mathcal{P}D  A_{\beta}}&+\braket{ \mathcal{P}A_{\alpha}}{(\delta D)  A_{\beta}}\\&\hspace{-0.6cm}+\braket{ \mathcal{P}A_{\alpha}}{D \delta A_{\beta}} \big).
\end{align}
Defining
$ R_{g}^{T}=r^{ji}\ip{g^{-1}T_{i}g}{\bullet} g^{-1}T_{j}g$ and $D^{T}=\frac{1}{ 1-\eta R^{T}_{g} \mathcal{P} } $
allows us to use the identities
\begin{align}
\label{eq:15}
 & \ \delta D =\eta D (\delta R_{g})\mathcal{P} D,\\
 & \ip{\delta R_{g} X}{ Y}=\ip{\left[ R_{g}X, \epsilon  \right]}{Y}+\ip{ [R_{g}^{T}  Y,\epsilon]}{X},
\end{align}
to write the variation as
\begin{align}
\label{eq:YBactionvariation}
\delta_{g}S=\int d^{2}\sigma   \bigg[\braket{ \epsilon}{ \partial_{\alpha}\mathcal{P}I^{\alpha}}+\braket{ [I_{\alpha} ,\epsilon]}{ \mathcal{P}I^{\alpha}}  \bigg],
\end{align}
where $I^{\alpha}=D A_{-}^{\alpha}+D^{T}A_{+}^{\alpha}$, matching the conventions of \cite{vanTongeren:2018vpb}. The equations of motion can now be identified using the symmetric space grading and remaining adjoint invariance to be
\begin{align}
\label{eq:YBeom}
\mathcal{E} \equiv \partial_{\alpha}\mathcal{P}I^{\alpha}-[I_{\alpha},\mathcal{P} I^{\alpha}]=0,
\end{align}
These equations are formally identical to those of the undeformed model, replacing $I$ by $A$.

\subsection{Homogeneous deformations}

For semi-simple Yang-Baxter deformations, the curvature of the deformed current
\begin{equation}
\curvJ\equiv \partial_{\alpha}I^{\alpha}_{+}-\partial_{\alpha} I^{\alpha}_{-}+[I_{\alpha},I^{\alpha}_{-}],
\end{equation}
is important in finding a Lax connection for the deformed model. In fact, for homogeneous semi-simple deformations the deformed current is flat on shell \cite{Kawaguchi:2014qwa,Matsumoto:2015jja}, so that at this level the model is formally identical to the undeformed model. This means the equations of motion admit the standard Lax representation
\begin{align}
\label{eq:laxansatz}
  L(z)&=I^{(0)}+\ell_{-}(z) I^{(1)}_{-}+ \ell_{+}(z) I^{(1)}_{+},
\end{align}
with the spectral parameter $z$ entering via $\ell_\pm = z^{\pm1}$. We hope to find the same in our non-semi-simple setting.

Given the lack of full adjoint invariance, we will work in components. Using flatness of $A$ expressed in terms of $I$, for arbitrary $T \in \mathfrak{g}$ we find
\begin{align}
\label{eq:defcurv}
  &\ip{T}{\curvJ}=c_{1}\eta+c_{2}\eta^{2},\\\nonumber
c_{1}&=\frac12\ip{T}{ (R_{g}^{T}-R_{g})\mathcal{E}}+ \frac12 \ip{T}{(R_{g}^{T}+R_{g})\mathcal{P} \mathcal{F}(I)}\\&\hspace{2.7cm}-\ip{\left[ \left( R_{g}^{T}+R_{g} \right)T, \mathcal{P}I^{\alpha}_{-} \right]}{\mathcal{P} I_{\alpha}},
  \\\nonumber
   c_{2}&=\ip{T}{[ R_{g}^{T}\mathcal{P} I_{\alpha},R_{g}\mathcal{P} I^{\alpha}_{-}]}+\ip{[R_{g}^{T} \mathcal{P}  I_{\alpha},R_{g}^{T} T]}{ \mathcal{P} I^{\alpha}_{-}}\\&\hspace{3.75cm}+\ip{[ R_{g} T,R_{g} \mathcal{P} I_{-}^{\alpha}]}{ \mathcal{P} I_{\alpha}}.
\end{align}
The leading order term vanishes on shell provided $R_g$ is antisymmetric, $R_{g}^{T}=-R_{g}$, while the second order term vanishes if
\begin{align}
\label{eq:RgsortofCYBE}
\ip{[R_{g} X,R_{g}Y]}{Z}+\ip{[R_{g} Z,R_{g}X]}{Y\hspace{-1pt}}+\ip{[R_{g} Y,R_{g} Z]}{X\hspace{-1pt}}=0,
\end{align}
for all $X,Y\in \mathfrak{g}^{(1)}$ and $Z\in \mathfrak{g}$. This equation is implied by
\begin{align}
\label{eq:nonsemisimpleoperatorCYBE}
\ip{[R X,R Y]}{Z}+\ip{[R Z,R X]}{Y\hspace{-1pt}}+\ip{[R Y,R Z]}{X\hspace{-1pt}}=0,
\end{align}
for $X,Y,Z \in \mathfrak{g}$, where $R$ is $R_g$ with $g$ the identity, i.e. as in equation \eqref{eq:Rnog}.\footnote{At least for the Poincar\'e algebra and its counterparts with different signature, these two equations are manifestly equivalent. Considering the coset representative $g= e^{x^\mu p_\mu}$, the various $x$-dependent contributions to equation \eqref{eq:RgsortofCYBE}, imply the ``missing'' components of \eqref{eq:nonsemisimpleoperatorCYBE}. Since local Lorentz transformations reorganize the $x$ dependence in an invertible fashion, this analysis is independent of the choice of coset representative.}

Equation \eqref{eq:nonsemisimpleoperatorCYBE} is the operator form of the classical Yang-Baxter equation in our setting. Starting from the standard matrix CYBE \eqref{eq:matrixCYBE} with $c=0$, it is nothing but $\ip{[[r,r]]}{X\otimes Y \otimes Z} = 0$. Conjugating $g^{-1}$ in each space, $\ip{[[r_g,r_g]]}{X\otimes Y \otimes Z} = 0$ equivalently gives equation \eqref{eq:RgsortofCYBE} without restrictions on its arguments. In short, provided $R$ is antisymmetric and solves the CYBE \eqref{eq:nonsemisimpleoperatorCYBE}, or equivalently $r$ satisfies the usual CYBE, also in our non-semi-simple setting the equations of motion admit the Lax representation \eqref{eq:laxansatz}. As in the undeformed setting discussed below equation \eqref{eq:undeflax}, for the Poincar\'e algebra we strictly speaking only need $\ell_+\neq \ell_-$ as opposed to $\ell_+ \ell_- = 1$ for deformations of general symmetric spaces.

\subsection{Inhomogeneous deformations}

We can also find a Lax connection for inhomogeneous deformations of the flat space string, although this requires additional structure. In essence this is because the tensor Casimir for the Poincar\'{e} algebra,
\begin{equation}\label{eq:PoincareOmega}
\Omega=3 \left(m_{\mu\nu}\wedge p^{\mu} \wedge p^{\nu}\right),
\end{equation}
has nonvanishing
\begin{equation}\label{eq:PoincareOmega}
\ip{\Omega}{p_\mu,p_\nu}_{1,2} = m_{\mu\nu},
\end{equation}
while the Poincar\'e algebra does not have a matching commutator ($[p_\mu,p_\nu]=0$).

Concretely, for inhomogeneous deformations the deformed current is no longer flat at second order in the deformation parameter. Assuming that the $R$ operator solves the inhomogeneous CYBE the curvature of the deformed current in \eqref{eq:defcurv} becomes
\begin{align}
\label{eq:inhomcurv}
\curvJ =-\eta R_{g}\mathcal{E}+\eta^{2}c^{2}  \ip{p^{\mu}}{ I^{\alpha}_{+}}\ip{p^{\nu}}{ I_{\alpha}}m_{\mu\nu}.
\end{align}
In the semi-simple case, the Lax ansatz \eqref{eq:laxansatz} works also for inhomogeneous deformations, by making the parameters $\ell_{+}$ and $\ell_{-}$ deformation parameter ($\eta$) dependent to compensate for the extra term in the curvature of the deformed current. This does not work in our case since commutators of grade one elements are all zero and can hence never cancel a grade zero component in \eqref{eq:inhomcurv}.

One way to get around this, is to define our Lax connection over a different algebra than our model.  From the general structure and the derivation in the homogeneous case, it is clear that if we modify our algebra such that the commutator $[p_\mu,p_\nu]$ is proportional to $m_{\mu\nu}$ while the other commutators are unchanged, our ansatz is likely to work. Fortunately, this can be readily achieved by replacing $\mathfrak{iso}(1,d-1)$ by $\mathfrak{so}(2,d-1)$ with
\begin{equation}\label{eq:so(2,d-1)comm}
[p_\mu,p_\nu] =  m_{\mu\nu}
\end{equation}
or by $\mathfrak{so}(1,d)$ with
\begin{equation}\label{eq:so(1,d)comm}
[p_\mu,p_\nu] = -m_{\mu\nu}
\end{equation}
and other commutators as in the Poincar\'e algebra. Concretely, the ansatz \eqref{eq:laxansatz} with its Poincar\'{e} generators replaced by those of $\mathfrak{so}(2,d-1)$ or $\mathfrak{so}(1,d)$ results in\footnote{The modification does not affect the equations of motion $\mathcal{E}$ that appear in the curvature of the Lax connection, since the grade zero components of $\mathcal{E}$ vanish identically, regardless of the underlying algebra.}
\begin{align}
\curvL =\frac{1}{2}( \ell_{+}&-\ell_{-} ) \mathcal{E} -\eta R_{g}\mathcal{E} +\frac{1}{2} (2-\ell_{+}-\ell_{-})\eta\mathcal{P}R_{g}\mathcal{E} \nonumber \\&+ (\eta^{2}c^{2}  \pm  \ell_{+}\ell_{-}) \ip{p^{\mu}}{I^{\alpha}_{+}}\ip{p^{\nu}}{I_{\alpha}}m_{\mu\nu} ,
\end{align}
where the upper sign applies for $\mathfrak{so}(2,d\hspace{-0.02cm}-\hspace{-0.02cm}1)$, and the lower for $\mathfrak{so}(1,d)$. We see that for $\mathfrak{so}(2,d\hspace{-0.005cm}-\hspace{-0.005cm}1)$ we get a spectral parameter dependent Lax connection for e.g. $\ell_{\pm} = i \eta c  z^{\pm1}$, while for $\mathfrak{so}(1,d)$ we can take $\ell_{\pm} =\eta c  z^{\pm1}$.\footnote{With this choice of $\ell_\pm$, $\mathfrak{so}(2,d\hspace{-0.02cm}-\hspace{-0.02cm}1)$ gives a real Lax connection for real $z$ for nonsplit deformations, while $\mathfrak{so}(1,d)$ does the same for split solutions, although of course we are always free to e.g. multiply $\ell_\pm$ by $\pm i$ anyway.} To take a nontrivial homogeneous or undeformed limit we can keep e.g. $\tilde{z} = \eta c z$ as our finite spectral parameter.\footnote{This does not give the standard symmetric space sigma model Lax connection, or the one for homogeneous deformations, but still one fitting the relevant constraints discussed below equation \eqref{eq:undeflax}.}

\section{Deformations of $\mathbb{R}^{1,2}$}

To illustrate the kind of deformations these $R$ operators generate, we will consider $\mathbb{R}^{1,2}$ as the simplest insightful case.\footnote{Up to automorphisms, $\mathbb{R}^{1,1}$ admits only the well-known $r$ matrices $p_0 \wedge p_1$ (abelian), $m_{01} \wedge p_+$ (jordanian), $m_{01} \wedge p_{0}$ (nonsplit), and $m_{01} \wedge p_{1}$ (split). The homogeneous cases are moreover trivial geometrically, see footnote 26 in \cite{Hoare:2016hwh}.} Here we have a variety of nontrivial deformations, which have moreover been classified \cite{Stachura:1998}, see also \cite{Kowalski-Glikman:2019ttm}. Of course, any homogeneous deformation can be immediately extended to a deformation of $\mathbb{R}^{1,d-1}$, in particular the bosonic critical string. Interestingly the same applies to inhomogeneous deformations, since we can freely add flat directions corresponding to trivially integrable massless scalar fields. Beyond this, three dimensional strings are also interesting in their own right. They admit a light-cone gauge quantization that preserves the Poincar\'e algebra \cite{Mezincescu:2010yp,Mezincescu:2011nh}, although recent results indicate that there is still an anomaly \cite{Dubovsky:2021cor}, now global.

Before discussing particular deformations, we note that Yang-Baxter deformations in general break isometries, and, if present, can break Weyl-invariance. The manifest symmetries of a Yang-Baxter model are the symmetries of the $r$ matrix, generated by those $T\in\mathfrak{g}$ for which $(1 \otimes \ad_{T} + \ad_{T} \otimes 1) r = 0$ with corresponding Noether current $j^{\alpha}=\ip{g^{-1}Tg}{\mathcal{P} I^{\alpha}}$. Moreover, as in the semisimple case, our deformations are guaranteed to preserve one-loop Weyl invariance if the associated $r$ matrix is unimodular \cite{Borsato:2016ose,Hronek:2020skb}.\footnote{Unimodularity is equivalent to replacing wedge products in an $r$ matrix by commutators, and finding zero.} In our setting this follows from the geometry based analysis of \cite{Hronek:2020skb}, because the deformed geometries still follow from the general formula $G+B=\left( G_0^{-1} + r \right)^{-1}$, where $G$ and $B$ are the deformed metric and $B$ field, $G_0$ is the undeformed metric (here Minkowski), and $r^{\mu\nu}=r^{ij}\chi_{i}^{\mu}\chi_{j}^{\nu}$ where $\chi_{i}$ is the Killing vector corresponding to the generator $T_i$ in a given set of coordinates \cite{Borsato:2018idb}, see also e.g. \cite{Sakamoto:2017cpu}. Other geometry-based results on e.g. the relation between homogeneous Yang-Baxter deformations and TsT transformations \cite{Osten:2016dvf} and nonabelian T duality \cite{Hoare:2016wsk,Borsato:2016pas,Borsato:2018idb}, similarly directly apply.

The homogeneous CYBE on $\mathfrak{iso}(1,2)$  has many inequivalent solutions. We focus on the three that are both nonabelian and unimodular \cite{Borsato:2016ose}
\begin{equation}\label{eq:homogeneousrmatrices}
\begin{aligned}
 r_{1}&=p_{0}\wedge m_{12} +\kappa  \left(p_{1} \wedge p_{2} \right),\\
  r_{2}&= p_{1} \wedge m_{02}+ \kappa \left( p_{0} \wedge p_{2} \right),\\
  r_{3}&= \frac{1}{4}\left( p_{0}+p_{2} \right)\wedge \left( m_{01}-m_{12} \right) + \kappa  \left(p_{1} \wedge p_{2} \right),
\end{aligned}
\end{equation}
with $\kappa$ as a free real parameter.
In terms of symmetries, these preserve
\begin{equation}\label{eq:homogeneousrmatrixsymmetries}
\begin{aligned}
  r_{1}&: p_{0}, m_{12},&
  r_{2}&:p_{1},m_{02} ,&
  r_{3}&:p_{1},(p_{0}+p_{2}).
\end{aligned}
\end{equation}
To manifest these symmetries, we can use the coset representatives
\begin{equation}
\begin{aligned}
  g_{1}&=e^{m_{12} \theta} e^{p_{0} t}e^{p_{1}r},\\
   g_{2}&=e^{m_{02} \theta} e^{p_{0} \rho}e^{p_{1}x},\\
  g_{3}&=e^{p_{1}x}e^{\left( p_{0}-p_{2} \right) x_{-}}e^{\left( p_{0}+p_{2} \right)x_{+}}.
\end{aligned}
\end{equation}
The corresponding deformed line elements and B fields are
\begin{equation}\label{eq:homogeneousbackgrounds}
  \begin{aligned}
     ds_{1}^{2}&= \frac{r^{2}d\theta^{2}-\eta^{2} \left( \kappa dr +r dt \right)^{2}}{1-\eta^{2}\left( r^{2}-\kappa^{2} \right)}-dt^{2}+dr^{2},\\
  ds^{2}_{2}&=\frac{\rho^{2}d\theta^{2}-\eta^{2} \left( \kappa d\rho -\rho dx \right)^{2}}{1+\eta^{2}\left( \rho^{2}-\kappa^{2} \right)}-d\rho^{2}+dx^{2},\\
  ds_{3}^{2}&=\frac{dx^{2}-\eta^{2} \left( x_{-}dx_{-}-\kappa dx_{+} \right)^{2}}{1+ \eta^{2}\kappa x_{-}}-4dx_{+}dx_{-},\\
   B_{1}&=\frac{r\eta}{1-\eta^{2}\left( r^{2}-\kappa^{2} \right)}  d\theta\wedge \left( \kappa d r+r dt   \right)\\
   B_{2}&=\frac{\rho\eta}{1+\eta^{2}\left( \rho^{2}-\kappa^{2} \right)} d\theta \wedge \left(  \kappa d \rho-\rho dx  \right)\\
   B_{3}&= \frac{\eta }{1+\eta^{2}\kappa  x_{-}} dx \wedge \left(\kappa dx_{+}-  x_{-} dx_{-}   \right),
\end{aligned}
\end{equation}
where, here and below, $\eta$ is our deformation parameter, and for the third background we shifted $x^- \rightarrow x^- -\kappa$.

Next, to illustrate inhomogeneous deformations, we consider all four inequivalent solutions to the inhomogeneous CYBE on $\mathfrak{iso}(1,2)$ \cite{Stachura:1998}
\begin{equation}\label{eq:inhomogeneousrmatrices}
\begin{aligned}
  r_{4}&= m_{0\mu}\wedge p^\mu, &  r_{5}&= m_{1\mu}\wedge p^\mu,\\
  r_{6}&= r_5 + m_{02}\wedge p_{1},& r_{7}&= \frac{1}{2} \epsilon^{\mu\nu\rho}m_{\mu\nu}\wedge p_{\rho},
\end{aligned}
\end{equation}
where $\epsilon$ is totally antisymmetric with $\epsilon^{012}=1$. Here we omit extra abelian terms made out of symmetries of the $r$ matrices, which can always be added. $r_{4}$ is the only nonsplit solution ($c=i$), the remaining are split solutions ($c=1$). $r_4$ and $r_5$ are the well-known timelike and spacelike $\kappa$-Poincar\'{e} $r$ matrices \cite{Lukierski:1993wxa} respectively. Interestingly, $r_{7}$ is unimodular, meaning it gives a purely bosonic inhomogeneous deformation that can be completed to a bosonic Weyl-invariant string background in 26 dimensions, at least to one loop. The preserved symmetries are
\begin{equation}\label{eq:inhomogeneousrmatrixsymmetries}
\begin{aligned}
 r_{4}&: m_{12}, p_{0},&
   r_{5}&: m_{02}, p_{1},\\
  r_{6}&: m_{02}, p_{1}, (p_{0}+p_{2}),&
  r_{7}&: m_{01},m_{02},m_{12}.
\end{aligned}
\end{equation}
We see that $r_7$ is also special in this regard, since it preserves a nonabelian $\mathfrak{so}(1,2)$.\footnote{Suggestively, this means that in this case the Yang-Baxter action with $R_g$ replaced by $R$ still has $H$ gauge symmetry. However, the analogue of equation \eqref{eq:YBactionvariation} then has $[A,\epsilon]$ in place of $[I,\epsilon]$, and the symmetric space structure no longer suffices to bring the equations of motion to a form analogous to equation \eqref{eq:YBeom}, meaning it is not obviously an integrable deformation.} The form of $r_7$ suggests it is a three dimensional peculiarity. The difference between $r_5$ and $r_6$ is abelian, but we include it since its presence enhances symmetry -- we are free to add $\alpha m_{02} \wedge p_1$ to $r_5$, but only for $\alpha=1$ we find enhanced symmetry.\footnote{For completeness, to $r_6$ we can also add the \emph{nonabelian} term $\beta m_{02}\wedge (p_0+p_2)$. This breaks the symmetries to those generated by $p_1$ and $p_0+p_2$, giving a different class of deformation. For brevity we do not present the resulting two parameter deformed metric.}

Here we can consider the coset representatives
\begin{equation}
\begin{aligned}
  g_{4}&=g_{1},&
  g_{5}&= g_6 =g_{2},&
  g_{7}&=e^{ m_{02}\phi} e^{m_{01}\theta}e^{ p_{0}\rho},\\
\end{aligned}
\end{equation}
with deformed line elements
\begin{equation}
\begin{aligned}
  ds^{2}_{4}&=\frac{-dt^{2}+dr^{2}}{1-r^{2} \eta^{2}}+r^{2} d\theta^{2},\\
  ds^{2}_{5}&=\frac{-d\rho^{2}+dx^{2}}{1-\rho^{2} \eta^{2}}+\rho^{2} d\theta^{2},\\
  ds^{2}_{6}&=-\eta^{2}\rho^{2}\left( d\rho+\rho d\theta \right)^{2}- d\rho^{2}+\rho^{2}d\theta^{2}+ dx^{2},\\
  ds^{2}_{7}&=-d\rho^{2}+ \frac{\rho^{2} d\theta^{2}+\rho^{2}\cosh^{2}(\theta) d\phi^{2}}{1+ \rho^{2}\eta^{2}},
\end{aligned}
\end{equation}
where for $r_7$ we scaled the deformation parameter by 1/2. The corresponding B fields are
\begin{equation}
\begin{aligned}
  B_{4}&=\frac{r\eta}{1-r^{2} \eta^{2}} dr  \wedge dt,&
  B_{5}&=\frac{\rho \eta}{1-\rho^{2}\eta^{2}} d\rho \wedge dx,\\
  B_{6}&= \eta \rho(d\rho +\rho d\theta) \wedge dx ,&
  B_{7}&= \frac{  \rho^{3}\eta\cosh(\theta)}{1+\rho^{2}\eta^{2}} d\theta\wedge d\phi.
\end{aligned}
\end{equation}
As expected \cite{Arutyunov:2014cra,Pachol:2015mfa}, the fourth and fifth background are T dual to $\mathrm{dS}_3$ (in $t$) and $\mathrm{AdS}_3$ (in $x$) respectively. The sixth background is also T dual to $\mathrm{AdS}_3$ (in $x$). This is natural because $r_5$ and $r_6$ differ by an abelian term containing $p_1$, so that the backgrounds are related by a TsT transformation involving $x$ \cite{Osten:2016dvf}, and hence are diffeomorphic in the $T$ dual frame. The sixth background is actually $\mathrm{AdS}_2 \times \mathbb{R}$, although the $B$ field mixes the two spaces.\footnote{In terms of Cartesian light-cone coordinates, it takes the simple form $ds_6^2 = - \eta^{2 }x_{-}^{2} dx_{+}^{2}-2 dx_{+}dx_{-}+ dx^{2}$, $B_6 = \eta x_- dx_+ \wedge dx$.}

\section{Conclusions and outlook}

We extended Yang-Baxter deformations to non-semi-simple symmetric space sigma models, in particular the flat space bosonic string. This requires a simple modification of the action, which does not actually affect the action in the semi-simple case.\footnote{A similar construction does not exist for the principal chiral ($G$) model, which always requires an $\mathrm{Ad}_G$ invariant bilinear form. In our language, viewing the principal chiral model as a symmetric space model ($(G\times G)/G$), gauge invariance requires an $\mathrm{Ad}_G$ invariant bilinear form on $\mathfrak{g}$.} In our setting, the classical Yang-Baxter equation for the $R$ operator can in general no longer be simplified to a freestanding operator equation, although of course the structure for the underlying $r$ matrix is unchanged. For homogeneous deformations the Lax connection can be constructed as in the semi-simple setting, while for inhomogeneous deformations this construction requires replacing the underlying non-semi-simple algebra by a suitable semi-simple one -- for $\mathfrak{iso}(n,d-n)$ we can take $\mathfrak{so}(n+1,d-n)$ or $\mathfrak{so}(n,d-n+1)$. We illustrated our discussion through nonabelian deformations of $\mathbb{R}^{1,2}$. Interestingly this includes an inhomogeneous deformation that is unimodular, and can be used to give a bosonic string background in $D=3$ or $D=26$.

There are various natural open questions. For one, we are now in a position to systematically investigate particular deformations of the bosonic flat space string, where it might be possible to get explicit insight into their quantum structure. We hope to be able to report on this in the near future. Furthermore, it would be interesting to extend our considerations to semi-symmetric space sigma models, i.e. the flat space Green-Schwarz string \cite{Henneaux:1984mh}. Here we can expect to lift certain nonunimodular deformations to unimodular ones by suitable fermionic extensions, as in the semi-simple setting \cite{Hoare:2018ngg,vanTongeren:2019dlq}. In general, given the existence of at least one bosonic inhomogeneous but unimodular deformation, it would also be insightful to investigate the new types of deformations permitted by non-semi-simple symmetry in both the bosonic and the supersymmetric setting. In terms of our three dimensional examples in particular, the three dimensional string can be quantized in light-cone gauge, but unfortunately all nonabelian deformations break the requisite isometries. Nevertheless it may be interesting to investigate these, and deformation of the $D=3$ superstring, in more detail.

From a light-cone gauge perspective, the nonsplit inhomogeneous deformation $(r_4)$ stands out since it preserves time translation symmetry, and therefore would admit a standard light-cone gauge if we add a further flat direction, e.g. in $D=26$. Moreover, the flat space $\kappa$-Poincar\'{e} deformation arises as a maximal deformation limit of the standard nonsplit deformation of anti-de Sitter space \cite{Arutyunov:2014cra,Pachol:2015mfa}. As such, we may expect the scattering theory of this model to be related to a subsector of the $\mathrm{AdS}_3$ one, although of course we need to keep in mind the lack of unimodularity of this deformation. It is of course also possible to consider unimodular nonabelian deformations of Euclidean space -- take e.g. $r = \frac{1}{2} \epsilon^{ijk} m_{ij} \wedge p_k$ in three dimensions -- and add a timelike and a suitable number of spatial directions, to get unimodular models that admit a light-cone gauge.

\section*{Acknowledgements}

We would like to thank Riccardo Borsato and Tomasz Trze\'{s}niewski for helpful discussions and Riccardo Borsato for comments on the draft. This work is supported by the German Research Foundation (DFG) via the Emmy Noether program ``Exact Results in Extended Holography''. ST is supported by LT.

\bibliographystyle{elsarticle-num}

\bibliography{Stijnsbibfile}


\end{document}